\documentclass[fleqn,twoside]{article}
\usepackage{espcrc2}

\newcommand{\plb}[3]{Phys. Lett. {\bf B#1} (#2) #3} 

\newcommand{\prd}[3]{Phys. Rev. {\bf D#1} (#2) #3}
\newcommand{\npb}[3]{Nucl. Phys. {\bf B#1} (#2) #3}
\newcommand{\npbps}[3]{Nucl. Phys. {\bf B}(Proc. Suppl.) {\bf #1} (#2) #3}

\newcommand{\jhep}[3]{JHEP {\bf #1} (#2) #3}

\newcommand{\be}{\begin{eqnarray}}
\newcommand{\ee}{\end{eqnarray}}
\newcommand{\bc}{\begin{center}}
\newcommand{\ec}{\end{center}}

\newcommand{\mathfrak}{\cal}

\def\non{\nonumber}

\def\g5{\gamma_5}

\newcommand{\AmS}{{\protect\the\textfont2
  A\kern-.1667em\lower.5ex\hbox{M}\kern-.125emS}}

\title{CP breaking in lattice chiral gauge theory}
\author{Kazuo Fujikawa\address[MCSD]
{Department of Physics, University of Tokyo, Bunkyo-ku, Tokyo 113, Japan},
Masato Ishibashi\addressmark\thanks{Talk presented by M. Ishibashi}
 and Hiroshi Suzuki\address{Department of Mathematical Sciences, Ibaraki 
University, Mito 310-8512, Japan}}
\begin{document}
\begin{abstract}
The CP symmetry is not manifestly implemented for the local and 
doubler-free Ginsparg-Wilson operator in lattice chiral gauge theory.
We precisely identify where the effects of this CP breaking 
appear.
\vspace{1pc}
\end{abstract}
\maketitle
\section{Introduction}
Discovery of gauge covariant local lattice Dirac
operators~\cite{Hasenfratz:1998ft,Neuberger:1998fp}, which satisfy the
Ginsparg-Wilson relation~\cite{Ginsparg:1982bj}, paved a way to a
manifestly local and gauge invariant lattice formulation of anomaly-free
chiral gauge theories~\cite{Luscher:1999du}--\cite{Kikukawa:2001kd}.
(See also related early
work in ref.~\cite{Narayanan:1993wx})

It has been however pointed out that the CP symmetry, the fundamental
discrete symmetry in chiral gauge theories, is not manifestly
implemented for the fermion action in this 
formulation, using the conventional 
Ginsparg-Wilson relation~\cite{Hasenfratz:2001bz}(This holds for 
general GW 
operators~\cite{Fujikawa:2000my}, as 
long as they are local and free of species 
doublers~\cite{Fujikawa:2002is}).
We calculate the fermion generating functional and identify where 
the effects of this CP breaking appear in this 
formulation~\cite{Fujikawa:2002is2}.

\section{Formulation}
In this section we calculate the fermion generating 
functional, which is given by
\begin{eqnarray}
\label{original}
&& Z_{\rm F}^{\{v,\overline v\}}[U,\eta,\overline\eta;t]
   =\int{\rm D}[\psi]{\rm D}[\overline\psi]\,e^{-S_{\rm F}},
 \\  
&&S_{\rm F}=a^4\sum_x[\overline\psi(x)D\psi(x)-
\overline\psi(x)\eta(x)-\overline\eta(x)\psi(x)]\non
\end{eqnarray}
where $\eta(x)$ and $\bar{\eta}(x)$ are source fields.
First, we introduce a one-parameter family of lattice analog of 
$\gamma_5$:
\begin{eqnarray}
  && \gamma_5^{(t)}
={\gamma_5-ta\gamma_5 D\over\sqrt{1+t(t-2)a^2D^\dagger D}},
\end{eqnarray}
where $D$ satisfies the conventional Ginsparg-Wilson relation:
$\gamma_5 D + D\gamma_5 =2aD\gamma_5 D$~\footnote{The following
analyses are valid for a more general form of the Ginsparg-Wilson relation
\cite{Fujikawa:2002is2}.} and  $\gamma_5^{(t)}$
satisfies $(\gamma_5^{(t)})^2=1$.
The ``conjugate'' of $\gamma_5^{(t)}$ is defined by
\begin{eqnarray}
&& \overline\gamma_5^{(t)}=\gamma_5\gamma_5^{(2-t)}\gamma_5.
\end{eqnarray}
Using $\gamma_5^{(t)}$ and its conjugate, chiral projection operators
are written as
\begin{eqnarray}
  && P_\pm^{(t)}={1\over2}(1\pm\gamma_5^{(t)}),\quad
   \overline P_\pm^{(t)}={1\over2}(1\pm\overline\gamma_5^{(t)}).
\end{eqnarray}
Then the chirality may be defined by \footnote{For definiteness, we
consider the left-handed Weyl fermion.}
\begin{eqnarray}
 && P_-^{(t)}\psi=\psi,\quad\overline\psi\overline P_+^{(t)}
=\overline\psi.
\end{eqnarray}

To define the fermion integration measure,
we introduce an ideal basis $v_j$ and $\bar{v}_k$
(which means that the formulation has 
locality, smoothness and gauge invariance when this basis is
used~\footnote{Refer to ref.\cite{Luscher:1999du} in further details.}) as
follows
\begin{eqnarray}
  && P_-^{(t)}v_j(x)=v_j(x),\quad
   \overline v_k\overline P_+^{(t)}(x)=\overline v_k(x),
\end{eqnarray}
and expand the fields as
\begin{eqnarray}
  && \psi(x)=\sum_jv_j(x)c_j,\quad
   \overline\psi(x)=\sum_k\overline c_k\overline v_k(x).
\end{eqnarray}
Then the integration measure is defined by
\begin{eqnarray}
  && {\rm D}[\psi]{\rm D}[\overline\psi]
   =\prod_j{\rm d}c_j\prod_k{\rm d}\overline c_k.
\end{eqnarray}

We next introduce a convenient basis $\{w,\overline w\}$ to
evaluate the fermion generating functional $Z_{\rm F}^{\{v,\overline v\}}$.  
For two different choices of basis, 
$\{v,\overline v\}$ and~$\{w,\overline w\}$, we have
\begin{eqnarray}
  && Z_{\rm F}^{\{v,\overline v\}}[U,\eta,\overline\eta;t]
   =e^{i\theta[U;t]}
   Z_{\rm F}^{\{w,\overline w\}}[U,\eta,\overline\eta;t].
\end{eqnarray}
Under $\delta_\xi U(x,\mu)=a\xi_\mu(x)U(x,\mu)$,
the variation of the phase is given by 
\begin{eqnarray}
  && \delta_\xi\theta[U;t]
   =-{\mathfrak L}_\xi^{\{v,\overline v\}}[U;t]
   +{\mathfrak L}_\xi^{\{w,\overline w\}}[U;t],
\end{eqnarray}
where the ``measure term'' is defined by
\begin{eqnarray}
&&   {\mathfrak L}_\xi^{\{v,\overline v\}}[U;t]
   =i\sum_j(v_j,\delta_\xi v_j)
   +i\sum_k(\delta_\xi\overline v_k^\dagger,
\overline v_k^\dagger).\non\\
\end{eqnarray}
We define an auxiliary basis $w$ as
\begin{eqnarray}
  && D^\dagger Dw_j(x)
   ={\lambda_j^2\over a^2}w_j(x),
\qquad
   \lambda_j\geq0
\non\\
  && w_j(x)=P_-^{(t)}w_j(x),\qquad(w_j,w_k)=\delta_{jk}.
\end{eqnarray}
Solving the eigenvalue problem of $D^\dagger D$, we can
classify $w_j$ as follows,
\begin{eqnarray}
   &&\mbox{i)}\lambda_j=0, \qquad w_{0\alpha_-}^-(x)
\nonumber\\
   &&\mbox{ii)}\lambda_j\neq 0,1,\qquad w_j^-(x)
\nonumber\\
   &&\mbox{iii)}\lambda_j=1,\qquad
\cases{\Psi_{+\beta_+}(x),& for $t>1$ \cr
            \Psi_{-\beta_-}(x),& for $t<1$ \cr}\non
\end{eqnarray}  
where $\alpha_-=1,\cdots, n_-,\quad \beta_+=1,\cdots, N_+$ and
$\beta_-=1,\cdots, N_-$. For $\bar{w}_k$, we 
adopt the left eigenfunctions of 
$DD^\dagger$:
\begin{eqnarray}
&& \mbox{i)}\lambda_j=0,\quad w_{0\alpha_+}^{+\dagger}(x)
\non\\
&&\mbox{ii)}\lambda_j\neq 0, \quad
\overline w_j(x)={a\over\lambda_j}w_j^\dagger D^\dagger(x)
\non
\end{eqnarray}
where $\alpha_+=1,\cdots, n_+$. Once having specified
basis vectors $\{w,\overline w\}$, it is straightforward
to perform the integration in $Z_{\rm F}^{\{w,\overline w\}}$. 
After some calculation, we have
\begin{eqnarray}
   &&Z_{\rm F}^{\{v,\overline v\}}[U,\eta,\overline\eta;t]
   =e^{i\theta[U;t]}\left({ 1 \over a}\right)^N\non\\
&&
\times   
\prod_{\alpha_-=1}^{n_-}
\left[a^4\sum_x\overline\eta(x)w_{0\alpha_-}^-(x)\right]
   \non\\
&&\times\prod_{\alpha_+=1}^{n_+}
\left[a^4\sum_x w_{0\alpha_+}^{+\dagger}(x)\eta(x)
\right]
   \nonumber\\
&&
\times\prod_{\lambda_j>0\atop\lambda_j\neq 1}
   \left({\lambda_j\over a}\right)
   \exp\left[a^8\sum_{x,y}\overline\eta(x)G^{(t)}(x,y)
   \eta(y)\right],
\non\\
\end{eqnarray}
where  
\begin{eqnarray}
&&N=\cases{N_+,&for $t>1$\cr
            N_-,&for $t<1$\cr},\non
\end{eqnarray}
and the propagator $G^{(t)}$ has been defined by
\begin{eqnarray}
  && DG^{(t)}(x,y)=\overline P_+^{(t)}(x,y)
   -\sum^{n_+}_{\alpha_+=1}w_{0\alpha_+}^+(x)w_{0\alpha_+}^{+\dagger}(y).\non
\end{eqnarray}

\section{CP transformed generating functional}
We adopt the standard CP transformation:
\begin{eqnarray}
  &&\psi(x)\to-W^{-1}\overline\psi^T(\bar x),\qquad
   \overline\psi(x)\to\psi^T(\bar x)W
\nonumber\\
   &&U(x,\mu)\to U^{\rm CP}(x,\mu)=\cases{
   {U(\bar x-a\hat i,i)^{-1}}^* \cr
   U(\bar x,4)^* \cr},\nonumber
\end{eqnarray}
where $W=\gamma_2$ and $\bar{x}=(-x_i,x_4),\quad (i=1,2,3)$. 
Let us consider the CP transformed generating functional
\begin{eqnarray}
\label{cptrans}  
  && Z_{\rm F}[U^{\rm CP},-W^{-1}\overline\eta^T,\eta^TW;t]
   =\int{\rm D}[\psi]{\rm D}[\overline\psi]\,e^{-S_{\rm F}}
\non\\
  && S_{\rm F}=a^4\sum_x[\overline\psi(x)D(U^{\rm CP})\psi(x)
   \\
&&\qquad\qquad\qquad +\overline\psi(x)W^{-1}\overline\eta^T(x)
   -\eta^T(x)W\psi(x)]
\non
\end{eqnarray}
Now introducing the new basis vectors
\begin{eqnarray}
&& v_k'=(\overline v_kW^{-1})^T,\qquad
   \overline v_j'=(-Wv_j)^T,\non
\end{eqnarray}
in eq.(\ref{cptrans}), 
we can show that
\begin{eqnarray}
 Z_{\rm F}[U^{\rm CP},-W^{-1}\overline\eta^T,\eta^TW;t]
   =Z_{\rm F}[U,\eta,\overline\eta;2-t].\non
\end{eqnarray}
Thus the sole effect of the CP transformation is given by
the change of parameter, $t\to 2-t$.
This change causes the exchange of $\Psi_+$ and~$\Psi_-$ and thus 
$N\to\overline N$, where
\begin{eqnarray}
 &&  \overline N=\cases{N_-&for $t>1$,\cr
            N_+&for $t<1$.\cr}\non
\end{eqnarray} 

In conclusion, the relation between the CP transformed generating
functional~(\ref{cptrans}) and the original 
generating functional~(\ref{original}) is
written as~\footnote{One can show that measure 
terms ${\mathfrak L}_\xi^{\{w,\overline w\}}$ and 
${\mathfrak L}_\xi^{\{v,\overline v\}}$ are invariant
under~$t\to2-t$~\cite{Fujikawa:2002is2}, which means
\begin{eqnarray}
  \delta_\xi\theta[U;t]
&=&\delta_\xi\theta[U;2-t].\non
\end{eqnarray}
As a result,
\begin{eqnarray}
&& \theta[U;2-t]-\theta[U;t]=\theta_M\non
\end{eqnarray}
where the constant $\theta_M$ is assigned for each topological 
sector $M,\> U\in M$.
}
\begin{eqnarray}
   &&Z_{\rm F}[U^{\rm CP},-W^{-1}\overline\eta^T,\eta^TW;t]
\nonumber\\
   &&=Z_{\rm F}[U,\eta,\overline\eta;t]
   \times e^{i\theta_M}\left({1 \over a}\right)^{\overline N-N}\non\\
   &&\times
   {\exp\left[a^8\sum_{x,y}\overline\eta(x)G^{(2-t)}(x,y)
   \eta(y)\right]\over
   \exp\left[a^8\sum_{x,y}\overline\eta(x)G^{(t)}(x,y)
   \eta(y)\right]}.\non
\end{eqnarray}
From the above equation, we see that the CP breaking in this 
formulation appears in three places for
the pure chiral gauge theory:
(I)~Difference in the overall
constant phase~$\theta_M$.
(II)~Difference in the overall
coefficient~$(1/a)^{\overline N-N}$. 
(III)~Difference in the
propagator appearing in the external 
fermion lines, $G^{(t)}$ and $G^{(2-t)}$.
We discuss their implications in this order:
(I) and (II) may be absorbed into a redefinition
of the topological overall 
factor~\footnote{Note that the topological overall factor, with
which the topological sector are summed, is not fixed within
this formulation~\cite{Luscher:1999du}.}.
(III) It seems impossible to remedy $G^{(t)}\neq G^{(2-t)}$(For $t=1$,
$\gamma^{(1)}_5$ is singular for $a^2 D^\dagger D\simeq 1$.
Therefore we cannot adopt $t=1$.).
But the CP breaking for $t\neq 1$ is quite modest.
In particular, for the conventional choice, $t=2$,
\begin{eqnarray}
&& G^{(2-t)}(x,y)
   =G^{(t)}(x,y)-a\gamma_5{1\over a^4}\delta_{x,y},\non
\end{eqnarray} 
and the breaking appears as an contact term.
It is thus expected that this breaking is safely removed in a
suitable continuum limit in the case of pure chiral gauge theory.

\section{With Yukawa couplings}
To add the Yukawa coupling to the present formulation,
we introduce the right-handed Weyl fermion and the Higgs field.
By the same arguments as for the pure chiral gauge theory,
 for the perturbative treatment of Yukawa couplings we have
 the following result: i) When the Higgs has no vacuum expectation
 value(VEV), the situation is the same 
as for the pure chiral gauge theory.
ii) When the Higgs acquires VEV, a new situation arises. The difference 
in the
propagator becomes a non-local function on the lattice.

\section{Discussion}

In the presence of the Higgs VEV, the CP breaking effect becomes 
intrinsically non-local.  This non-local breaking could be serious 
in the non-perturbative treatment of the Higgs mechanics. As a 
related issue, the definition of Majorana fermions has certain 
complications in Ginsparg-Wilson operators~\cite{Fujikawa:2002i}.




\begin{thebibliography}{99}

\bibitem{Hasenfratz:1998ft}
P. Hasenfratz, \npbps{63}{1998}{53};
\npb{525}{1998}{401}.

\bibitem{Neuberger:1998fp}
H. Neuberger,
\plb{417}{1998}{141};
\plb{427}{1998}{353}.

\bibitem{Ginsparg:1982bj}
P.H. Ginsparg and K.G. Wilson,
\prd{25}{1982}{2649}.

\bibitem{Luscher:1999du}
M. L\"uscher,
\npb{549}{1999}{295};\npb{568}{2000}{162}.

\bibitem{Suzuki:1999}
H. Suzuki,
Prog.Theor.Phys. {\bf 101}(1999)1147

\bibitem{Suzuki:2000ii}
H. Suzuki,
\npb{585}{2000}{471}.

\bibitem{Luscher:2000zd}
M. L\"uscher,
\jhep{06}{2000}{028}.

\bibitem{Kikukawa:2001kd}
Y. Kikukawa and Y. Nakayama,
\npb{597}{2001}{519}.

\bibitem{Narayanan:1993wx}
R. Narayanan and H. Neuberger,
\npb{443}{1995}{305}.

\bibitem{Hasenfratz:2001bz}
P.~Hasenfratz,
\npbps{106}{2002}{159}.

\bibitem{Fujikawa:2000my}
K. Fujikawa,
\npb{589}{2000}{487}.

\bibitem{Fujikawa:2002is}
K.~Fujikawa, M.~Ishibashi and H.~Suzuki,
\plb{538}{2002}{197}

\bibitem{Fujikawa:2002is2}
K.~Fujikawa, M.~Ishibashi and H.~Suzuki,
\jhep{04}{2002}{046}

\bibitem{Fujikawa:2002i}
K. Fujikawa and M. Ishibashi, 
\npb{622}{2002}{115};
\plb{528}{2002}{295}.


\end{thebibliography}
\end{document}